# Crystal Structure of the Sodium Cobaltate Deuterate Superconductor $Na_xCoO_2 \cdot 4xD_2O$ ($x \approx 1/3$)

J. D. Jorgensen, M. Avdeev, D. G. Hinks, J. C. Burley, S. Short,
Materials Science Division, Argonne National Laboratory, Argonne IL  60439

ABSTRACT

Neutron and x-ray powder diffraction have been used to investigate the crystal structures of a sample of the newly-discovered superconducting sodium cobaltate deuterate compound with composition $Na_{0.31(3)}CoO_2 \cdot 1.25(2)D_2O$ and its anhydrous parent compound $Na_{0.61(1)}CoO_2$.  The anhydrous parent compound $Na_{0.61(1)}CoO_2$ has two partially occupied Na sites sandwiched, in the same plane, between $CoO_2$ layers.  When Na is removed to make the superconducting composition, the Na site that experiences the strongest Na-Co repulsion is emptied while the occupancy of the other Na site is reduced to about one third.  The deuterate superconducting compound is formed by coordinating four $D_2O$ molecules (two above and two below) to each remaining Na ion in a way that gives Na-O distances nearly equal to those in the parent compound.  One deuteron of the $D_2O$ molecule is hydrogen bonded to an oxygen atom in the $CoO_2$ plane and the oxygen atom and the second deuteron of each $D_2O$ molecule lie approximately in a plane between the Na layer and the $CoO_2$ layers.  This coordination of Na by four $D_2O$ molecules leads in a straightforward way to ordering of the Na ions and $D_2O$ molecules consistent with the observation of additional shorter-range scattering features in the diffraction data.  The sample studied here, which has $T_c$=4.5 K, has a refined composition of $Na_{0.31(3)}CoO_2 \cdot 1.25(2)D_2O$, in agreement with the expected 1:4 ratio of Na to $D_2O$.  These results show that the optimal superconducting composition should be viewed as a specific hydrated compound, not a solid solution of Na and $D_2O$ ($H_2O$) in $Na_xCoO_2 \cdot yD_2O$.  The hydrated superconducting compound may be stable over a limited range of Na and $D_2O$ concentration, but studies of $T_c$ and other physical properties vs. Na or $D_2O$ composition should be viewed with caution until it is verified that the compound remains in the same phase over the composition range of the study.

INTRODUCTION

The recent discovery of superconductivity at 4.5 K in hydrated sodium cobaltate with a reported composition of $Na_{0.3}CoO_2 \cdot 1.4H_2O$ **[1]** has generated renewed interest in the compositional chemistry and crystal structure of these materials. In recent years, the related anhydrous compound, $Na_xCoO_2$, has been studied because of its interesting transport properties. Wang et al. **[2]** have argued that spin entropy arising from the two-dimensional triangular Co lattice is the explanation for the enhanced thermopower.**[3]** More recently, the same authors have reported an anomalous Hall effect for $Na_xCoO_2$.**[4]** They observe that $Na_xCoO_2$ exhibits "strange metal physics as observed in the cuprates." $Na_xCoO_2$ is made superconducting by removing part of the Na, which changes the oxidation state of Co, and intercalating water to dramatically increase the separation between the $CoO_2$ layers. Bulk superconductivity with $T_c > 2$ K has been reported to exist over a very narrow range of Na compositions, approximately $1/4 \leq x \leq 1/3$, in the hydrated compound $Na_xCoO_2 \cdot 1.3H_2O$ with the maximum $T_c$ (4.5 K) at $x \approx 0.3$.**[5]** This dome-like behavior of $T_c$ is reminiscent of the layered cuprates, suggesting that this new layered compound may be the first non-copper example of the same physics. The importance of the two-dimensional nature of the structure is emphasized by the fact that a lower hydrate, $Na_{0.3}CoO_2 \cdot 0.6H_2O$, with the same Co formal oxidation state but substantially less separation between the $CoO_2$ layers is not superconducting.**[6]**

Prior to the recent interest in transport properties and superconductivity, compounds of this family were studied because of their fast-ion conducting properties.**[7]** The Na ions are in two partially occupied crystallographic sites, both of which lie in a plane between layers of edge-sharing $CoO_6$ octahedra. The ability to change the Na (or other cation) content and to intercalate water has been investigated for isostructural compounds such as $Na_{2/3}(Co_xNi_{1/3-x}Mn_{2/3})O_2$,**[8]** and $Li_{0.12}K_{0.35}MnO_{2.14} \cdot 0.45H_2O$.**[9]** However, the structural details of the water intercalation and, in particular, the crystal structure of the $Na_xCoO_2 \cdot yH_2O$, have never been reported. Clearly, an understanding of the crystal structure is fundamental to interpreting the physical behavior.

In this paper, we report the crystal structures of the starting anhydrous compound, $Na_{0.61}CoO_2$ and the deuterated superconducting composition, $Na_{0.31}CoO_2 \cdot 1.25D_2O$. We show that, when the Na concentration is reduced, Na is selectively removed from one of the two inequivalent Na sites. Then, after $D_2O$ is incorporated, the $D_2O$ molecules occupy sites between the $CoO_2$ layers and the Na ions in such a way that the coordination of Na to the oxygen ions of the $D_2O$ molecules is similar to its previous coordination to oxygen ions in the $CoO_2$ layers in $Na_{0.61}CoO_2$, except that the Na to $D_2O$ ratio is four, resulting in two $D_2O$ molecules above the Na ion and two below. One deuteron of each $D_2O$ molecule is hydrogen bonded to an oxygen atom in the $CoO_2$ layer. The constraint of minimizing Na-D and D-D repulsions leads in a straightforward way to two-dimensional ordering of Na ions and $D_2O$ molecules. In the supercell structure, the second deuteron of each $D_2O$ molecule is hydrogen bonded to the oxygen atom of a neighboring $D_2O$ molecule to form hydrogen bonded zigzag D-O•••D-O•••D-O...

chains. The diffraction data are consistent with this ordering in the plane and also show that ordering along the c axis, with a shorter length scale, is present. A possible explanation is that both Na ions and $D_2O$ molecules are ordered in the plane, but $D_2O$ ordering has a much shorter coherence length than Na ordering along the *c* axis. The refined composition of the sample studied here, $Na_{0.31(3)}CoO_2 \cdot 1.25(2)D_2O$ manifests the ideal 1:4 ratio of Na to $D_2O$. The findings reported here shed important new light on studies of $T_c$ vs. composition **[5]** which show a dome-like behavior of $T_c$ and a recent theory **[10]** that predicts that the highest $T_c$ is achieved at a composition between two ordered structures, where Co charge ordering is disrupted. This work shows that the maximum $T_c$ is achieved in an ordered structure, but with a composition slightly off the ideal ordering composition of $Na_{1/3}$.

SYNTHESIS

A powder sample of the anhydrous compound with composition $Na_{0.61(1)}CoO_2$ was made by repeated heating and grinding of a mixture of $Na_2CO_3$ and $Co_3O_4$ in air at 850° C. An initial powder stoichiometry of $Na_{0.7}CoO_2$ resulted in a sample of composition $Na_{0.61(1)}CoO_2$ due to a slight loss of Na during the heating cycles. (All compositions stated in the paper are those determined by Rietveld refinement using neutron powder diffraction data.) The chemical oxidative deintercalation of Na from this material is generally done using $Br_2$ **[11]** or $I_2$ **[12]** in acetonitrile. For the neutron diffraction experiments, a fully deuterated sample was desired in order to avoid the large neutron incoherent scattering from hydrogen, which adds background to the diffraction data. In order to eliminate any possible hydrogen incorporation during the oxidative deintercalation, either from the acetonitrile or residual $H_2O$ in the acetonitrile, we used $Br_2$ in $D_2O$ as the oxidizing medium. About 10 wt% of $Na_{0.61}CoO_2$ powder in $D_2O$ with a 100 to 200% excess of $Br_2$ was shaken overnight in an air-tight pyrex tube. The resulting powder was then separated by either filtration or centrifuging from the $Br_2$ solution with as little exposure to the atmosphere as possible. After washing with $D_2O$ to remove excess $Br_2$ and NaBr the resulting material was kept in a container with 100% relative humidity of $D_2O$.

This "as-oxidized" deuterated material contained free $D_2O$, as evidenced by the appearance of peaks from ice in a neutron powder diffraction pattern taken at low temperature. In order to remove only the free $D_2O$ and none of the lattice $D_2O$, knowledge of the vapor pressure of $D_2O$ in equilibrium with the higher deuterate phase is required. Approximately 2g of the material was placed in a vacuum system and pumped to remove air. The $D_2O$ pressure over the hydrate was measured using a capacitance manometer as the $D_2O$ was volumetrically removed by cooling a calibrated glass bulb attached to the vacuum system to liquid nitrogen temperature. To allow the sample to come to equilibrium at each partial pressure of $D_2O$, this cryopumping was done in small steps -- typically a few minutes of pumping followed by equilibration times of up to several hours. All measurements were done at $23.0 \pm 0.2°$ C. **Fig. 1** shows the $D_2O$ pressure as a function of removed $D_2O$. The long plateau at 2.3-2.4 kPa pressure indicates that the sample is a mixture of free $D_2O$ and the higher deuterate phase (plus, perhaps, some $D_2O$ adsorbed on the walls of the vacuum system). In this particular experiment, ~1.6 cc of free $D_2O$ was removed before reaching the point where little additional free $D_2O$ remained and the $D_2O$ pressure dropped sharply with further pumping. The plateau at about 1 kPa indicates the transition to a two-phase deuterate mixture; i.e., a mixture of the higher deuterate phase and the lower deuterate phase reported to have the composition $Na_{0.3}CoO_2 \cdot 0.6D_2O$.[6] A single phase of the higher deuterate with a minimum amount of free $D_2O$ can be made by equilibrating at a $D_2O$ pressure just above 1 kPa. The sample used for the diffraction measurements was equilibrated at 1.2 kPa $D_2O$ pressure overnight.

For neutron diffraction, the sample was loaded into an indium-sealed vanadium can in a helium-filled glove bag with only about a ~45 sec. exposure to the dry helium atmosphere during loading. Once the sample is sealed in the can with helium, there is not enough free volume to allow a significant change in the $D_2O$ content of the sample. The diffraction pattern showed only the higher deuterate phase; none of the lower deuterate phase **[6]** (which is easily identified by its shorter *c* axis) was present, indicating no loss of $D_2O$ during sample loading. Additionally, no $D_2O$ ice peaks were seen at low temperature, indicating that no free $D_2O$ existed in this sample. To minimize handling of the sample, AC susceptibility measurements were not done until after the neutron diffraction measurements (described below). The AC susceptibility data are shown in **Fig. 2**. A superconducting transition with an onset at 4.5 K is seen. This equals the highest onset temperature seen in this material, confirming that the sample is at an optimal superconducting composition. The width of the transition is similar to that seen for optimally-doped samples in other studies,**[5]** confirming the quality of the sample. This post-diffraction measurement of $T_c$ confirms that the sample remained at the superconducting composition during the neutron diffraction measurements.

NEUTRON AND SYNCHROTRON X-RAY POWDER DIFFRACTION

Neutron powder diffraction data were collected on the Special Environment Powder Diffractometer **[13]** at the Intense Pulsed Neutron Source for powder samples of $Na_{0.61}CoO_2$ and superconducting $Na_{0.31}CoO_2 \cdot 1.25D_2O$. Data were collected at various temperatures between room temperature (~295 K) and 12 K with the samples sealed in vanadium cans along with helium exchange gas and cooled using a Displex closed-cycle helium refrigerator.

For the superconducting sample, following the neutron diffraction experiment, high-resolution synchrotron x-ray diffraction data were collected at room temperature at beam line 5-BMC (DND-CAT) at the Advanced Photon Source (Argonne National Laboratory). The sample was sealed in a 1mm dia. glass capillary to prevent $D_2O$ loss. The high-resolution x-ray data were used to look for any subtle distortions of the fundamental hexagonal unit cell and to further investigate additional diffraction peaks from shorter-range ordering (to be discussed later).

RIETVELD STRUCTURE REFINEMENTS

The neutron powder diffraction data for both samples were analyzed by the Rietveld method using the previously reported [1] $P6_3/mmc$ hexagonal space group. Rietveld refinement profiles are shown in **Fig. 3.** This structural model correctly indexes all of the sharp diffraction peaks for both compositions. For the higher deuterate superconducting composition, additional scattering resulting from ordering on shorter length scales was also seen. The most prominent additional scattering is a broad, modulated feature extending from about 2.8-2.5 Å. Somewhat sharper features of the additional scattering can most clearly be seen in the difference plot of **Fig. 3b**. These broadened supercell peaks were also seen in the high-resolution x-ray diffraction data, where the difference in peak widths, compared to peaks of the fundamental hexagonal cell, was much more obvious. The interpretation of this supercell scattering will be discussed later.

The refined structural parameters and selected bond lengths at various temperatures for $Na_{0.61}CoO_2$ are given in **Table I**. The structure is shown in **Fig. 4**. The $CoO_2$ layers are formed from edge-sharing $CoO_6$ octahedra, where the Co-O bond length is 1.908 Å (at room temperature). The oxygen atoms lie in two planes, above and below the plane of Co atoms. Na ions lie in two partially occupied sites in a plane halfway between the $CoO_2$ layers. The two Na sites differ in their coordination to the oxygen atoms in the $CoO_2$ layers, as shown in **Fig. 5**. The Na1 site is coordinated to six oxygen atoms that are bonded to six different Co atoms (three above and three below). The Na2 site is also coordinated to six oxygen atoms (three above and three below), but the sets of three oxygen atoms (above and below Na) are bonded to the same Co atoms. The average Na-O bond lengths are identical (by symmetry) for the two Na sites (2.379 Å at 12 K). It has previously been proposed **[14]** that slightly weaker bonding of the Na2 site results from Na-Co Coulomb repulsion, which is greater for this site when both first- and second-nearest neighbor Na-Co repulsions are considered. This leads to a smaller occupancy for the Na2 site compared to the Na1 site. The Na ions show displacements from their ideal sites at both room temperature and low temperature. This probably results from repulsion of randomly-located neighboring Na ions, locally violating the hexagonal symmetry. In Rietveld refinement, such displacements can be modeled with anisotropic temperature factors or by assigning the Na ions to displaced, partially occupied sites. The latter model has been used for the Na1 site in the refinements reported in **Table I**. For the Na2 site, the displacement was not large enough to statistically justify a displaced-site model. Because this detail of the structural model does not give any additional insight into the compound, all figures show only the average, high-symmetry, Na sites and average Na-O bond lengths (i.e., bond lengths to the ideal, high-symmetry site). The tables, however, list the Na1-O bond lengths resulting from the use of a displaced-site model for Na1.

Rietveld refinement of the $Na_{0.31}CoO_2 \cdot 1.25D_2O$ structure was approached by first using the same model as for the anhydrous compound, with different lattice parameters, and not including any $D_2O$ in the model. These refinements immediately showed that when Na is removed, it comes selectively from the more weakly bonded Na2 site, leaving that site empty within the accuracy of the refinement. In the final refinements, the occupancy of the Na2 site tended to go slightly negative (~ one esd) so it was set to zero. To achieve the x≈0.3 composition, part of the Na is also removed from the Na1 site (which was initially only partially occupied). In the deuterate compound, the displacement of the Na1 ion off its ideal site is even larger than for the anhydrous compound. Thus, in the refinement model, Na1 has been assigned to a partially occupied [$6h(2x,x,1/4)$] displaced site.

To locate the $D_2O$ molecules, Fourier difference maps were plotted. These maps showed scattering density suggesting that the oxygen atoms of the $D_2O$ molecules were located in layers between the $CoO_2$ layers and Na layers. Additionally, the maps showed two possible deuterium sites: one approximately in the same plane as the oxygen atoms and one closer to the $CoO_2$ layers. Fourier maps from room-temperature and low-temperature data were not dramatically different. Rietveld refinements were then attempted by adding $D_2O$ molecules in general positions [$24l(x,y,z)$] with rigid-body constraints being used to maintain the expected molecule geometry; i.e., a D-O distance of 0.99Å and a D-O-D angle of 109°. This constraint is thought to be justified based on the configurations seen for water molecules in a wide range of hydrated compounds and ices.[15] In this model, the refinement gives the best-fit position for the oxygen atom of the $D_2O$ molecule and the orientation of the molecule. The resulting refined structural parameters and selected bond lengths for the deuterate superconducting compound are given in **Table II**. This refinement model gives a composition of $Na_{0.31(3)}CoO_2 \cdot 1.25(2)D_2O$ (based on the 12 K data where correlations between temperature factors and site occupancies are the smallest). The refined 1:4 ratio of Na to $D_2O$ (within one esd) suggests that the synthesis method used here, including the careful adjustment of the $D_2O$ content to the conditions for the most stable phase, has produced a compound with four $D_2O$ molecules coordinated to each Na ion. While sixfold coordination is common for Na ions, fourfold coordination is observed in many oxides and hydrated compounds, including $Na_2O$.[16] The refined average Na-O distance (2.31 Å) is also consistent with that expected for a four-coordinated Na hydrate [16,17] and is remarkably similar to the Na-O distance for Na ions in water (2.4 Å), where the average coordination is 4.9.[18] (Note that when static displacement of the Na ions is considered, the resulting inequivalent Na-O distances vary from 2.18 to 2.42 Å.) The refined composition also falls within the range of values reported for the composition of the optimized superconducting compound.[1,5,6]

STRUCTURE MODEL FOR THE DEUTERATE PHASE

Refinement of this structural model in the fundamental hexagonal unit cell leads to partially-occupied symmetry-equivalent sites for $D_2O$ that cannot be simultaneously occupied in the actual structure because of impossibly close O-O distances, incorrect orientations of $D_2O$ with respect to Na (i.e., orientations in which D, rather than O, is oriented toward the Na ion), and energetically unfavorable D-D repulsions. A plausible picture for the actual structure can be readily constructed by considering logical rules for chemical bonding and coordination. All symmetry-equivalent $D_2O$ molecules have the same coordination to oxygen atoms in the $CoO_2$ plane. Hence, the actual structure can most easily be visualized by considering only one layer of $D_2O$ molecules and the associated Na ions. **Fig. 6a** shows all symmetry-equivalent Na and $D_2O$ sites in one of these layers. If all such sites were occupied (including the $D_2O$ layers above and below Na), the composition would be $NaCoO_2 \cdot 12D_2O$. However, it is straightforward to see that all of these sites cannot be simultaneously occupied. Three bonding/coordination rules can be applied to construct plausible structures from **Fig. 6a**. First, to minimize D-Na repulsion, $D_2O$ molecules strongly prefer to be oriented with the deuteron "pointing" away from, not toward, an occupied Na site. Second, impossibly close oxygen-oxygen distances cannot occur. Third, each Na ion should be coordinated by four $D_2O$ molecules (two above and two below) in agreement with the refined composition and consistent with known Na oxide and hydrate compounds.[16,17]

The application of these rules leads in a straightforward way to the ordered structure shown in **Fig. 6b**. (The interested reader can sequentially mark Na and $D_2O$ sites in **Fig. 6a** following the three rules above and see how this solution is obtained.) Both the Na ions and the $D_2O$ molecules are ordered into a two-dimensional supercell. This structure has 1/3 of the Na sites occupied with four oxygen atoms (two above and two below) coordinated to each Na ion, giving an ideal composition of $Na_{1/3}CoO_2 \cdot (4/3)D_2O$. The $D_2O$ molecules are arranged in a way that hydrogen bonds can form between neighboring molecules in the plane to give zigzag D-O•••D-O•••D-O... chains, as shown in **Fig. 6b**. In the basal plane, this supercell is rectangular with dimensions $3a$ x $\sqrt{3}a$. Note that there are three equivalent choices for the orientation of $D_2O$ molecules around each Na ion, leading to three equivalent directions for these chains. One would not expect the orthorhombic supercell to have lattice parameters that so accurately match the fundamental hexagonal cell unless the three equivalent choices for forming ordered layers occur with equal populations. Thus, if there is ordering (i.e., translational symmetry of a particular stacking sequence) along the c axis, the *c* axis must also be tripled. Hence, the smallest supercell that can manifest the Na and $D_2O$ ordering while preserving the hexagonal symmetry of the fundamental cell is orthorhombic with dimensions $3a$ x $\sqrt{3}a$ x $3c$.

This orthorhombic $3a$ x $\sqrt{3}a$ x $3c$ supercell can describe all of the major features of the scattering from shorter-range ordering seen in neutron and x-ray powder diffraction data, as shown in **Fig. 7**. The broad scattering feature extending from ~2.8 to 2.5 Å in the neutron diffraction data can be explained equivalently as a (30*l*) rod of scattering with its intensity strongly modulated along *l* or as scattering from a series of (30*l*) peaks [i.e., (300), (301), (302), (303), etc.] whose peak broadening increases with *l*, causing them to overlap (**Fig. 7a**). Maxima in the intensity are observed around the (303) and (309) peaks. The scattering from this 30*l* zone is characteristic of short-range ordering in which the coherence length along the *c* axis is shorter than in the basal plane. Since this scattering feature is observed much more strongly in neutron diffraction than in x-ray diffraction, one might conclude that it comes from short-range ordering of $D_2O$ molecules (because D has a comparatively large scattering cross section for neutrons). In the x-ray diffraction data, the (30.12) peak in this series, seen at 2.45 Å, is stronger than for neutron diffraction and is resolved from the nearby peaks from the fundamental hexagonal cell by the high resolution (**Fig. 7b**). The width of this supercell peak defines a coherence length of about 500 Å, which is longer than the out-of-plane coherence length implied by the neutron diffraction data for the (30*l*) series (**Fig. 7a**). Thus, taken together, the neutron and x-ray diffraction data suggest that there may be two different length scales for supercell ordering along the *c* axis. A possible explanation is that Na site occupancies and/or displacements (and perhaps corresponding distortions of the $CoO_2$ layers) are ordered over a ~500 Å length scale while $D_2O$ orientations are ordered over a shorter length scale. The supercell peak at 1.629 Å, which is very strong in neutron diffraction and is also seen in x-ray diffraction, can be indexed as a (030) peak (**Fig. 7c**). The width of this peak defines an in-plane coherence length of about 1000 Å, which is assumed to apply to both Na and $D_2O$ in-plane ordering. The supercell peak at ~1.41 Å (seen in the difference plot in **Fig. 3b**) can be indexed as the (060)/(330) peak and may have additional scattering from (06$l$)/(33$l$) peaks extending to smaller d spacings as for the (30$l$) series. Many other weaker features can also be indexed in the same supercell, but, because the supercell is so large, the assignments are not unique. A full interpretation of the scattering from ordering on shorter length scales will certainly come in the future from studies of single crystals, which are now reported to have been grown.**[19,20]** For that reason, and given the complexity resulting from the very large supercell involved, a more complete analysis of the scattering from shorter-range ordering has not been pursued using the present data.

The proposed structure model of the ordered superconducting deuterate compound $Na_xCoO_2 \cdot 4xD_2O$ (x≈1/3) is illustrated in **Fig. 8**. The diffraction experiment done here cannot differentiate between the cis (planar) and trans (twisted by 60°) arrangements of the pairs of $D_2O$ molecules above and below a particular Na ion; both arrangements are shown in **Fig. 8**. Based on O-O and D-D repulsions, the trans configuration would perhaps be energetically preferred. Local probes of the Na coordination will be needed to determine which configuration exists in the actual structure.

DISCUSSION

The structure shown in **Figs. 6b & 8** satisfies several criteria expected to be obeyed for a hydrate compound containing Na.**[15-18]** The Na ions are coordinated by four $D_2O$ molecules, which is a common coordination for oxygen around Na.**[16,17]** The average Na-O distance (2.31 Å) is typical of such compounds and is, additionally, close to the average Na-O distance in the anhydrous compound $Na_{0.61}CoO_2$ (2.41 Å). The $D_2O$ molecules are oriented to minimize Na-D repulsion and achieve an Na-D distance in the acceptable range.**[15]** All $D_2O$ molecules achieve positions that allow hydrogen bonding involving both deuterons (as shown in **Figs. 6b & 8**).**[15]** The hydrogen bonding in the plane between neighboring $D_2O$ molecules leads to the formation of zigzag D-O•••D-O•••D-O... chains. The bending angle and D-D distance (2.44 Å) of these chains is typical of hydrogen bonded systems in which the donor and acceptor atoms are of the same type.**[15]** These many examples of agreement with the literature for hydrate compounds lead to the conclusion that $Na_xCoO_2 \cdot 4xD_2O$ (x≈1/3) is a typical sodium hydrate (deuterate) compound. There are no exotic features of the structure, although the layering is somewhat uncommon.

The results presented here show that the highest onset $T_c$ (=4.5 K) reported in $Na_xCoO_2 \cdot yD_2O$ is achieved in an ordered deuterate structure in which each Na ion is coordinated to four $D_2O$ molecules and both Na ions and $D_2O$ molecules are ordered into a superlattice. This ordered structure has the ideal composition $Na_xCoO_2 \cdot 4xD_2O$ with x=1/3 when all Na and $D_2O$ sites are occupied. The refined composition, $Na_{0.31(3)}CoO_2 \cdot 1.25(2)D_2O$, suggests that the maximum $T_c$ is achieved at a composition slightly off the ideal composition, in agreement with other reports for the composition at maximum $T_c$.**[1,5,6]** In the sample studied here, which was made in a way that places the composition at a plateau in partial pressure of $D_2O$ in equilibrium with the sample, the ideal 1:4 Na to $D_2O$ ratio is maintained, even though the sample is slightly deficient in both Na and $D_2O$ compared to the ideal model. Other reports **[1,5,6,19]** indicate that a higher water content can be achieved for an Na content of $x \approx 0.3$. It is perhaps possible that additional water could be accommodated using the space made available by Na vacancies for the $x \approx 0.3$ composition (see **Fig. 6b**). One can visualize a hydrogen-bonded network of $D_2O$ molecules occupying the available space in a way consistent with the pattern of symmetry equivalent positions shown in **Fig. 6a**. Alternatively, additional $D_2O$ could perhaps be accommodated in the Na plane, as in the structure of the lower deuterate phase, $Na_{0.3}CoO_2 \cdot 0.6D_2O$.**[21]** Thus, reports of water contents higher than 4/3 are not necessarily inconsistent with the model presented here, especially in the case where the Na content is below 1/3. However, the incorporation of two water molecules per formula unit, as reported by Jin et al. **[19]** seems unlikely in light of the structure. Perhaps reports of water concentrations higher than 4/3 should be viewed with suspicion until it is shown that all of the water is actually in the lattice. Surface adsorbed water could easily cause errors in bulk measurements of water content. The data presented here for the partial pressure of $D_2O$ in equilibrium with the hydrated material (**Fig. 1**) argue that the water content does not depart far from the ideal 1:4 ratio of Na to $D_2O$ molecules. Even if extra water can be accommodated in the space made available by Na vacancies, as the ideal Na content of 1/3 is approached the ability to accommodate extra water would be expected to decrease, Additionally, for Na contents greater than 1/3, the compound must either generate "Na interstitial" defects in the Na ordering, where the extra Na ions could not achieve the fourfold coordination to $D_2O$ molecules, or phase separate. The data of Schaak et al. **[5]** for $T_c$ vs. Na concentration are consistent with phase separation for Na contents above $x=1/3$. $T_c$ remains constant at ~2 K, while the superconducting fraction decreases, for increasing Na contents above 1/3.

It is important to speculate about what happens for Na concentrations below $x \approx 0.3$. Baskaran **[10]** has presented a theory based on the idea that the highest $T_c$ in this system occurs between the compositions $Na_{1/3}$ and $Na_{1/4}$ that would be expected to give ordered phases. One hypothesis of his theory is that ordering of Na ions would promote charge ordering of the Co atoms, which, in turn, would suppress superconductivity. Thus, the ordered phases at the ideal compositions $x=1/3$ and 1/4 would be unfavorable for superconductivity. In agreement with these ideas, the maximum $T_c$ is achieved at a Na concentration below $x=1/3$.**[5]** However, the present results show that the optimal superconducting composition displays ordering, over an in-plane coherence length of ~1000 Å, the same as for the ideal $x=1/3$ composition. The increase of $T_c$ upon decreasing the Na content from $x=1/3$ could simply be a doping effect, as proposed by Schaak et al.**[5]** To better understand this behavior, it is important to understand what happens when the Na concentration is further reduced toward the composition $x=1/4$, which corresponds to a different ordered structure in Baskaran's model. **Fig. 6c** shows a possible ordered structure with this Na concentration. At this Na concentration, each Na ion can achieve sixfold coordination to $D_2O$ molecules. This is also the most common coordination for Na oxide and hydrate structures.**[16]** The present data give no evidence for the occurrence of this ordered compound. However, if it does form at the Na composition $x=1/4$, the data of Schaak et al. for $T_c$ vs. composition may have a different interpretation. The drop in $T_c$ from 4.5 K to 2 K upon decreasing the Na concentration from 0.30 to 0.26 may result from a transition from the $Na_{1/3}$ to the $Na_{1/4}$ phase. If this is the case, the conclusion that the dome-like behavior of $T_c$ mimics the behavior seen in the layered cuprates **[5]** may be premature. Clearly, a careful structural study of a composition near x=1/4 must be pursued.

CONCLUSIONS

In conclusion, we have shown that the highest reported $T_c$ (=4.5 K) in the $Na_xCoO_2 \cdot yD_2O$ system is achieved in a hydrate phase with well-defined composition and ordering of both Na ions and $D_2O$ molecules. The deuteration of Na leads to the formation of $2D_2O$-Na-$2D_2O$ "pillars" extending between the $CoO_2$ planes. This interesting "pillar" structure is further stabilized by hydrogen bonding within the planes of the $D_2O$ molecules to form zigzag D-O•••D-O•••D-O... chains. These structural results are in contrast to initial reports that implied that the material could be viewed a solid solution that could intercalate both Na and $D_2O$ in varying amounts. The ideal composition of the superconducting phase is $Na_xCoO_2 \cdot 4xD_2O$ (x≈1/3). The sample studied here has a refined Na composition of x=0.31(3) and a perfect 1:4 Na to $D_2O$ ratio (within one esd) consistent with the ideal structure, but with vacancy defects on both the Na and $D_2O$ sites; i.e., entire $2D_2O$-Na-$2D_2O$ "pillar" sites are vacant. . This hydrated phase would be expected to exist only within rather narrow composition limits. For x≥1/3, fourfold coordination of Na by $D_2O$ molecules cannot be achieved. As x approaches 1/4, a different ordered phase, involving sixfold coordination of the Na, could be formed. Thus, measurements of physical properties, such as $T_c$, vs. composition should be viewed with caution until it is confirmed that all samples in a particular study are in the same phase.

ACKNOWLEDGEMENTS

This work is supported by the U. S. Department of Energy, Basic Energy Sciences - Materials Sciences, under contract No. W-31-109-ENG-38.

REFERENCES

1.  K. Takada, H. Sakurai, E. Takayama-Muromachi, F. Izumi, R. A. Dilanian, T. Sasaki, Nature <u>422</u>, 53 (2003)

2.  Y. Wang, N. Rogado, R. J. Cava, N. P. Ong, Nature <u>423</u>, 425 (2003)

3.  I Terasaki, Y. Sasago, K. Uchinokkura, Phys. Rev. B <u>56</u>, R12685 (1997)

4.  Y. Wang, N. Rogado, R. J. Cava, N. P. Ong, cond-mat/0305455

5.  R. E. Schaak. T. Klimczuk, M. L. Foo, R. J. Cava, cond-mat/0305450

6.  M. L. Foo, R. E. Schaak, V. L. Miller, T. Klimczuk, N. S. Rogado, Y. Wang, G. C. Lau, C. Craley, H. W. Zandbergen, N. P. Ong, R. J. Cava, Solid State Commun $\underline{127}$, 33 (2003)

7.  See, for example, J.-J. Braconnier, C. Delmas, C. Fouassier, P. Haggenmuller, Mat. Res. Bull. $\underline{15}$, 1797 (1980)

8.  Z. Lu, J. R. Dahn, Chem. Mater. $\underline{13}$, 1252 (2001)

9.  S. H. Kim, W. M. Im, J. K. Hong, S. M. Oh, J. Electrochem. Soc. $\underline{147}$, 413 (2000)

10. G. Baskaran, cond-mat/0306569

11. S. Kikkawa, S. Miyazaki and M. Koizumi, J. Solid State Chem. $\underline{62}$, 35 (1986)

12. B. L. Cushing and J. B. Wiley, J. Solid State Chem. $\underline{141}$, 385 (1998)

13. J. D. Jorgensen, J. Faber, Jr., J. M. Carpenter, R. K. Crawford, J. R. Haumann, R. L. Hitterman, R. Kleb, G. E. Ostrowski, F. J. Rotella, and T. G. Worlton, J. Appl. Crystallogr. $\underline{22}$, 321 (1989)

14. R. J. Balsys, R. L. Davis, Solid State Ionics $\underline{93}$, 279 (1996); Y.-J. Shin, M.-H. Park, J.-H. Kwak, H. Namgoong, O. H. Han, Solid State Ionics $\underline{150}$, 363 (2002)

15. W. H. Baur, Acta Cryst. B$\underline{28}$, 1456 (1972); W. H. Baur, Acta Cryst. B$\underline{48}$, 745 (1992); G. Chiari and G. Ferraris, Acta Cryst. B$\underline{38}$, 2331 (1982); T. Steiner, Acta Cryst. B$\underline{54}$, 464 (1998)

16. R. D. Shannon and C. T. Prewitt, Acta Cryst. B$\underline{25}$, 925 (1969)

17. P. J. A. Wunderlich, Acta Cryst. $\underline{10}$, 462 (1957)

18. N. T. Skipper and G. W. Neilson, J. Phys.: Condens. Matter $\underline{1}$, 4141 (1989)

19. R. Jin, B. C. Sales, P. Khalifah, D. Mandrus, cond-mat/0306066

20. F. C. Choi, J. H. Cho, P. A. Lee, E. T. Abel, K. Matan, Y. S. Lee, cond-mat/0306659

21. M. Avdeev, J. D. Jorgensen, D. G. Hinks, J. Burley, S. Short, to be published

TABLES

Table I. Refined structural parameters for $Na_{0.61}CoO_2$ based on Rietveld refinement using neutron powder diffraction data for various temperatures. The structure is refined in hexagonal space group $P6_3/mmc$ with Co at $2a(0,0,0)$, O at $4f(1/3,2/3,z)$, Na1 at $6h(2x,x,1/4)$, and Na2 at $2b(0,0,1/4)$. The Na1 ion position is refined at the displaced $6h$ site, rather than its average, high-symmetry site $2d(1/3,2/3,3/4)$, in order to better model the static and dynamic displacements seen for this ion. Site occupancies are expressed in terms of the chemical formula unit to simplify understanding the refined composition.

| Temperature, K | 12 | 25 | 50 | 100 | 150 | 200 | 250 | 295 |
|---|---|---|---|---|---|---|---|---|
| a, Å | 2.83176(3) | 2.83176(4) | 2.83170(4) | 2.83160(4) | 2.83167(4) | 2.83179(4) | 2.83203(4) | 2.83287(2) |
| c, Å | 10.8431(2) | 10.8433(3) | 10.8442(3) | 10.8487(3) | 10.8573(3) | 10.8682(3) | 10.8810(3) | 10.8969(1) |
| z(O) | 0.09057(7) | 0.09060(10) | 0.0906(1) | 0.0904(1) | 0.0903(1) | 0.0903(1) | 0.0903(1) | 0.09024(5) |
| n(Na1) | 0.438(6) | 0.429(12) | 0.441(12) | 0.426(12) | 0.414(12) | 0.417(12) | 0.411(12) | 0.396(6) |
| 2x(Na1) | 0.577(2) | 0.576(2) | 0.580(3) | 0.570(2) | 0.572(2) | 0.578(3) | 0.575(3) | 0.574(2) |
| n(Na2) | 0.171(5) | 0.168(5) | 0.172(8) | 0.184(9) | 0.165(8) | 0.152(8) | 0.166(9) | 0.170(5) |
| $U_{11}=U_{22}=2U_{12}(Co)$, Å$^2$ | 0.0020(3) | 0.0018(4) | 0.0019(5) | 0.0018(5) | 0.0024(5) | 0.0018(5) | 0.0022(5) | 0.0030(3) |
| $U_{33}(Co)$, Å$^2$ | 0.0065(7) | 0.0070(10) | 0.0081(10) | 0.0064(10) | 0.0055(11) | 0.0072(11) | 0.0083(11) | 0.0072(6) |
| $U_{11}=U_{22}=2U_{12}(O)$, Å$^2$ | 0.0043(2) | 0.0043(2) | 0.0045(2) | 0.0044(2) | 0.0044(2) | 0.0054(2) | 0.0054(2) | 0.0061(1) |
| $U_{33}(O)$, Å$^2$ | 0.0084(2) | 0.0076(3) | 0.0088(4) | 0.0085(3) | 0.0096(4) | 0.0099(4) | 0.0105(4) | 0.0105(2) |
| U(Na1), Å$^2$ | 0.005(1) | 0.005(2) | 0.006(2) | 0.004(2) | 0.004(2) | 0.009(2) | 0.011(2) | 0.008(1) |
| U(Na2), Å$^2$ | 0.004(1) | 0.003(1) | 0.003(2) | 0.005(2) | 0.004(2) | 0.006(1) | 0.012(2) | 0.015(2) |
| Co-O (x6), Å | 1.9072(4) | 1.9068(5) | 1.9071(6) | 1.9064(5) | 1.9065(6) | 1.9070(6) | 1.9078(5) | 1.9084(3) |
| Na1-O, Å  x4<br>x2 | 2.313(1)<br>2.535(3) | 2.313(2)<br>2.538(5) | 2.315(2)<br>2.530(5) | 2.311(2)<br>2.549(4) | 2.314(2)<br>2.548(5) | 2.319(2)<br>2.538(6) | 2.319(2)<br>2.545(6) | 2.321(1)<br>2.550(3) |
| Na2 – O (x6), Å | 2.3794(5) | 2.3799(8) | 2.3796(8) | 2.3814(8) | 2.3828(8) | 2.3842(8) | 2.3858(8) | 2.3887(4) |

Table II. Refined structural parameters for $Na_{0.31}CoO_2 \cdot 1.25 D_2O$ based on Rietveld refinement using neutron powder diffraction data for various temperatures. The structure is refined in hexagonal space group $P6_3/mmc$ with Co at $2a(0,0,0)$, O at $4f(1/3,2/3,z)$ and Na1 at $6h(2x,x,1/4)$). The Na1 ion position is refined at the displaced $6h$ site, rather than its average, high-symmetry site $2d(1/3,2/3,3/4)$, in order to better model the static and dynamic displacements seen for this ion.. The $D_2O$ molecule, consisting of $O_w$, D1, and D1 all in general positions $24l(x, y, z)$, is refined using rigid body constraints with D-O distances of 0.99 Å and a D-O-D angle of 109°. Site occupancies are expressed in terms of the chemical formula unit to simplify understanding the refined composition.

| Temperature, K | 12 | 48 | 56 | 75 | 100 | 150 | 200 | 295 |
|---|---|---|---|---|---|---|---|---|
| a, Å | 2.81693(5) | 2.81705(6) | 2.81708(6) | 2.81718(7) | 2.81746(8) | 2.81806(8) | 2.81896(8) | 2.82166(5) |
| c, Å | 19.6449(6) | 19.6475(7) | 19.6491(7) | 19.6543(9) | 19.6633(9) | 19.6880(9) | 19.7125(9) | 19.7681(6) |
| z(O) | 0.0473(1) | 0.0472(1) | 0.0471(1) | 0.0470(1) | 0.0472(1) | 0.0470(1) | 0.0469(1) | 0.0469(1) |
| n(Na1) | 0.31(3) | 0.31(3) | 0.32(3) | 0.32(3) | 0.32(3) | 0.32(3) | 0.33(3) | 0.35(3) |
| 2x(Na1) | 0.52(2) | 0.57(4) | 0.55(2) | 0.59(4) | 0.59(4) | 0.59(4) | 0.556(15) | 0.525(5) |
| U(Co), Å$^2$ | 0.0001(6) | 0.0007(5) | 0.0009(5) | 0.0011(4) | 0.0013(6) | 0.0014(8) | 0.0014(8) | 0.0040(6) |
| $U_{11}=U_{22}=2U_{12}(O)$, Å$^2$ | 0.0027(2) | 0.0028(3) | 0.0029(3) | 0.0031(3) | 0.0033(3) | 0.0034(3) | 0.0037(4) | 0.0046(2) |
| $U_{33}(O)$, Å$^2$ | 0.0146(7) | 0.0142(7) | 0.0146(7) | 0.0146(9) | 0.0162(9) | 0.0162(9) | 0.0190(10) | 0.0226(8) |
| $x(O_w)$ | -0.260(6) | -0.265(5) | -0.265(5) | -0.264(6) | -0.258(6) | -0.272(7) | -0.257(6) | -0.263(4) |
| $y(O_w)$ | -0.091(7) | -0.116(7) | -0.118(6) | -0.121(7) | -0.126(10) | -0.106(9) | -0.058(7) | -0.073(9) |
| $z(O_w)$ | 0.1750(6) | 0.1774(4) | 0.1775(3) | 0.1771(4) | 0.1769(5) | 0.1768(4) | 0.1721(4) | 0.1706(2) |
| x(D1) | -0.479(6) | -0.446(4) | -0.446(4) | -0.438(4) | -0.452(5) | -0.428(5) | -0.446(5) | -0.421(3) |
| y(D1) | -0.204(10) | -0.258(8) | -0.260(5) | -0.262(9) | -0.275(13) | -0.242(10) | -0.153(8) | -0.159(10) |
| z(D1) | 0.1321(3) | 0.1324(4) | 0.1325(3) | 0.1319(4) | 0.1328(5) | 0.1306(4) | 0.1272(4) | 0.1240(3) |
| x(D2) | -0.098(4) | -0.126(7) | -0.125(6) | -0.129(8) | -0.142(12) | -0.109(8) | -0.087(5) | -0.112(7) |
| y(D2) | 0.312(6) | 0.287(7) | 0.285(6) | 0.280(8) | 0.271(13) | 0.300(9) | 0.345(7) | 0.325(9) |
| z(D2) | 0.1819(5) | 0.1801(4) | 0.1801(3) | 0.1803(5) | 0.1804(5) | 0.1786(4) | 0.1800(5) | 0.1801(3) |
| $n(D_2O)$ | 1.25(2) | 1.24(2) | 1.24(2) | 1.22(2) | 1.23(2) | 1.24(2) | 1.28(2) | 1.43(2) |
| $U_{11}(O_w)=U_{11}(D1)=U_{11}(D2)$ | 0.080(5) | 0.074(4) | 0.071(4) | 0.074(5) | 0.085(7) | 0.0071(5) | 0.074(4) | 0.105(6) |
| $U_{22}(O_w)=U_{22}(D1)=U_{22}(D2)$ | 0.12(1) | 0.093(8) | 0.088(8) | 0.085(9) | 0.107(9) | 0.10(1) | 0.14(1) | 0.24(2) |
| $U_{33}(O_w)=U_{33}(D1)=U_{33}(D2)$ | 0.012(2) | 0.015(2) | 0.013(2) | 0.017(2) | 0.016(3) | 0.013(3) | 0.011(2) | 0.019(2) |
| $U_{12}(O_w)=U_{12}(D1)=U_{12}(D2)$ | -0.08(6) | 0.017(7) | 0.013(7) | 0.017(8) | 0.025(9) | 0.016(8) | 0.037(8) | 0.10(1) |
| $U_{13}(O_w)=U_{13}(D1)=U_{13}(D2)$ | -0.027(3) | -0.023(2) | -0.021(2) | -0.018(3) | -0.012(4) | -0.018(3) | -0.020(3) | -0.031(3) |
| $U_{23}(O_w)=U_{23}(D1)=U_{23}(D2)$ | -0.019(4) | -0.002(4) | -0.002(3) | 0.004(5) | 0.005(5) | -0.0008(9) | -0.036(5) | -0.065(6) |
| U(Na1) | 0.014(4) | 0.020(4) | 0.022(4) | 0.023(5) | 0.021(6) | 0.023(5) | 0.023(8) | 0.018(5) |
| Co-O (x6), Å | 1.8727(10) | 1.8722(10) | 1.8710(10) | 1.8702(12) | 1.8723(12) | 1.8713(13) | 1.8719(4) | 1.8744(9) |

FIGURE CAPTIONS

Fig. 1. Pressure of $D_2O$ as a function of the amount of $D_2O$ removed by cryopumping a sample of $Na_{0.31(3)}CoO_2 \cdot yD_2O$ at room temperature (~23° C.), where the starting sample contained both lattice $D_2O$ and excess liquid $D_2O$. Using the sample mass and the results from neutron powder diffraction for the amount of $D_2O$ in a sample equilibrated at 1.2 kPa, the amount of removed $D_2O$ has been expressed in terms of $D_2O$ molecules per formula unit. The plateau at P≈1 kPa is the coexistence region for both stable deuterates (y≈4/3, 2/3).

Fig. 2. AC susceptibility vs. temperature for the sample of $Na_{0.31}CoO_2 \cdot 1.25D_2O$ used for neutron and x-ray diffraction measurements.

Fig. 3. Rietveld refinement profiles for (a) $Na_{0.61}CoO_2$ at 12 K and (b) $Na_{0.31}CoO_2 \cdot 1.25D_2O$ at 12 K. All data are refined in the same hexagonal space group $P6_3/mmc$. Crosses are the raw time-of-flight neutron powder diffraction data. The solid line is the calculated diffraction pattern. Tick marks indicate the positions of allowed reflections. A difference curve (observed minus calculated) is plotted at the bottom. Additional features in the pattern for $Na_{0.31}CoO_2 \cdot 1.25D_2O$, e.g., the broad scattering at 2.5-2.8 Å and the sharper features seen in the difference plot at about 1.4 Å and 1.6 Å, are from supercell ordering on shorter length scales (discussed in the text).

Fig. 4. Crystal structure of $Na_{0.61}CoO_2$ (hexagonal space group $P6_3/mmc$). Co atoms are octahedrally coordinated to six oxygen atoms. The $CoO_6$ octahedra share edges to form layers. Na ions are in two partially occupied sites, Na1 and Na2, that cannot be simultaneously occupied in the same region of the cell because of impossibly close Na1-Na2 distances. Average Na sites are shown, ignoring small static displacements for Na1 used in the refinement model (see **Table I**).

Fig. 5. Coordination of the Na1 and Na2 sites in $Na_{0.61}CoO_2$ to nearby oxygen atoms in the $CoO_2$ layers. Na1 is coordinated to six oxygen atoms (three above and three below), which are coordinated to six different Co atoms. Na2 is also coordinated to six oxygen atoms, but these oxygen atoms are associated with two Co atoms (one directly above and one directly below the Na ion). The average Na-O bond lengths (2.379 Å at 12 K) are the same for the two Na sites, but the Na-Co distances are significantly different. Atom symbols are the same as for **Fig. 4**. Note that the Na1-O bond length shown here is the average bond length, i.e., the bond length to the average Na1 site, rather than to the displaced site used in the refinement (**Table I**).

Fig. 6. (a) View of one layer of Na ions and $D_2O$ molecules in the structure of $Na_{0.31}CoO_2 \cdot 1.25D_2O$ showing all symmetry-equivalent sites, as determined from Rietveld refinement using neutron powder diffraction data. Note that all symmetry-equivalent oxygen atoms are at the same z coordinate even though the structure-drawing program shows one above the other for overlapping pairs. Shown at (b) and (c) are two two-dimensional ordered structures that result in a straightforward by applying logical rules about bonding and coordination (discussed in the text). The supercell at (b) has an ideal composition of $Na_{1/3}CoO_2 \cdot (4/3)D_2O$,

with fourfold coordination of Na by $D_2O$, that agrees with the 1:4 Na to $D_2O$ ratio from the Rietveld structure refinement. It can be expressed as a rectangular supercell of dimensions $3a$ x $\sqrt{3}a$. Hydrogen bonding in the plane links neighboring $D_2O$ molecules to form zigzag D-O•••D-O•••D-O... chains, as shown by the dotted lines. The supercell at (c) has an ideal composition of $Na_{1/4}CoO_2 \cdot (3/2)D_2O$ with sixfold coordination of Na by $D_2O$, . It can be expressed as a hexagonal supercell of dimensions $2a$ x $2a$. Atom symbols are the same as for **Fig. 4**.

Fig. 7. Major features of the scattering from ordering on shorter length scales that can be explained by a $3a$ x $\sqrt{3}a$ x $3c$ orthorhombic supercell. (a) The broad scattering extending from 2.8-2.5 Å can be indexed as a series of $(30l)$ peaks whose broadening increases with increasing $l$. The $l$ index is given below the tick marks that mark the positions of these peaks. Equivalently, this scattering can be viewed as a $(30l)$ rod that is strongly modulated as a function of l. The sharp feature below 2.5 Å is the (008) peak from the fundamental hexagonal cell shown for comparison. (b) The (30.12) peak (marked by the arrow and dotted line) is seen in high-resolution x-ray diffraction, but not in neutron diffraction. Its width defines a coherence length of ~500 Å. (c) The (030) peak (marked by the arrow) is very strong in neutron diffraction and is also seen in x-ray diffraction. Its width defines a coherence length of ~1000Å in the basal plane. In (b) and (c), the other sharper features are peaks from the fundamental hexagonal unit cell, plus small peaks from $Co_3O_4$ in the x-ray data. All allowed peaks of the fundamental hexagonal cell are marked and indexed as $(hkl)_h$. (Data shown here are all at room temperature; thus, d spacings are slightly different from those in **Fig. 3** for data at 12 K.)

Fig. 8. View of the structure of the hydrated compound $Na_xCoO_2 \cdot 4xD_2O$ ($x \approx 1/3$) showing two possible local configurations for the four $D_2O$ molecules coordinated to each Na ion. The $CoO_2$ layers are held apart by $2D_2O$-Na-$2D_2O$ "pillars" that are hydrogen bonded to oxygen atoms in the $CoO_2$ layers. The $D_2O$ molecules can assume either the cis (planar) (a) or the trans (twisted by 60°) arrangement (b). The view of the structure shown here is a partial unit cell that ignores the supercell ordering in which $2D_2O$-Na-$2D_2O$ "pillar" locations are only ~1/3 occupied according to the two-dimensional ordering pattern shown in **Fig. 6b**. Atom symbols are the same as for **Fig. 4**.

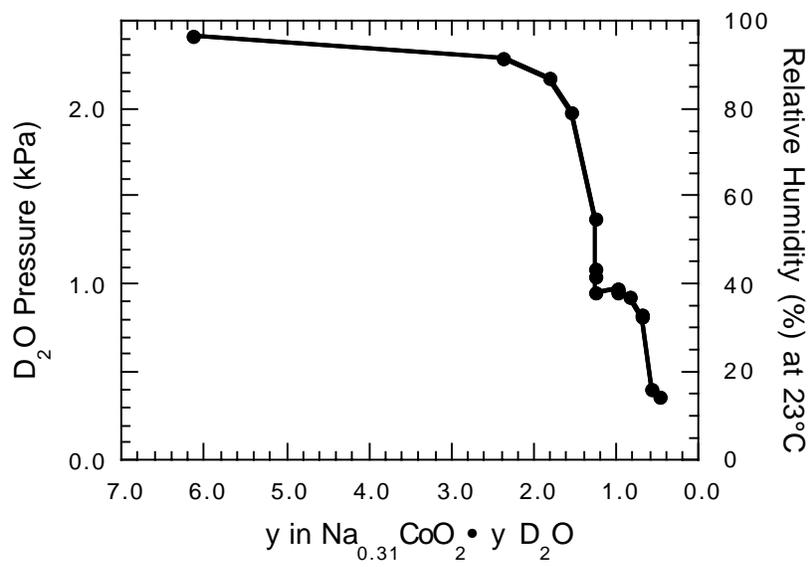

Fig. 1.

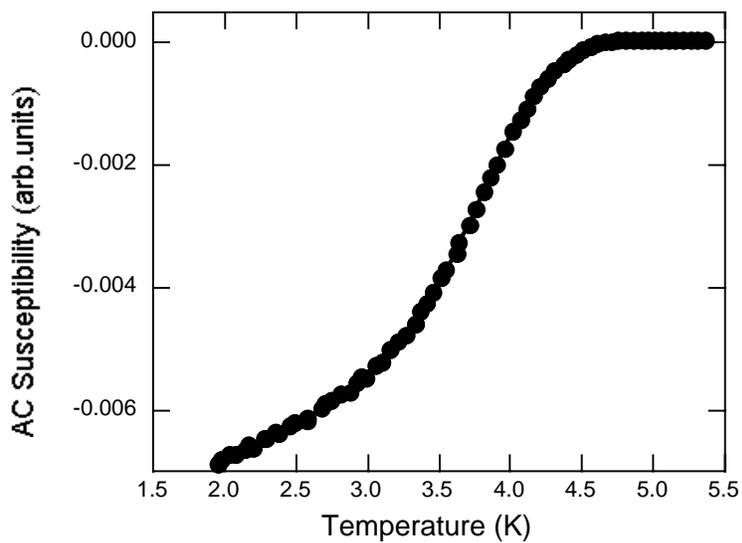

Fig. 2

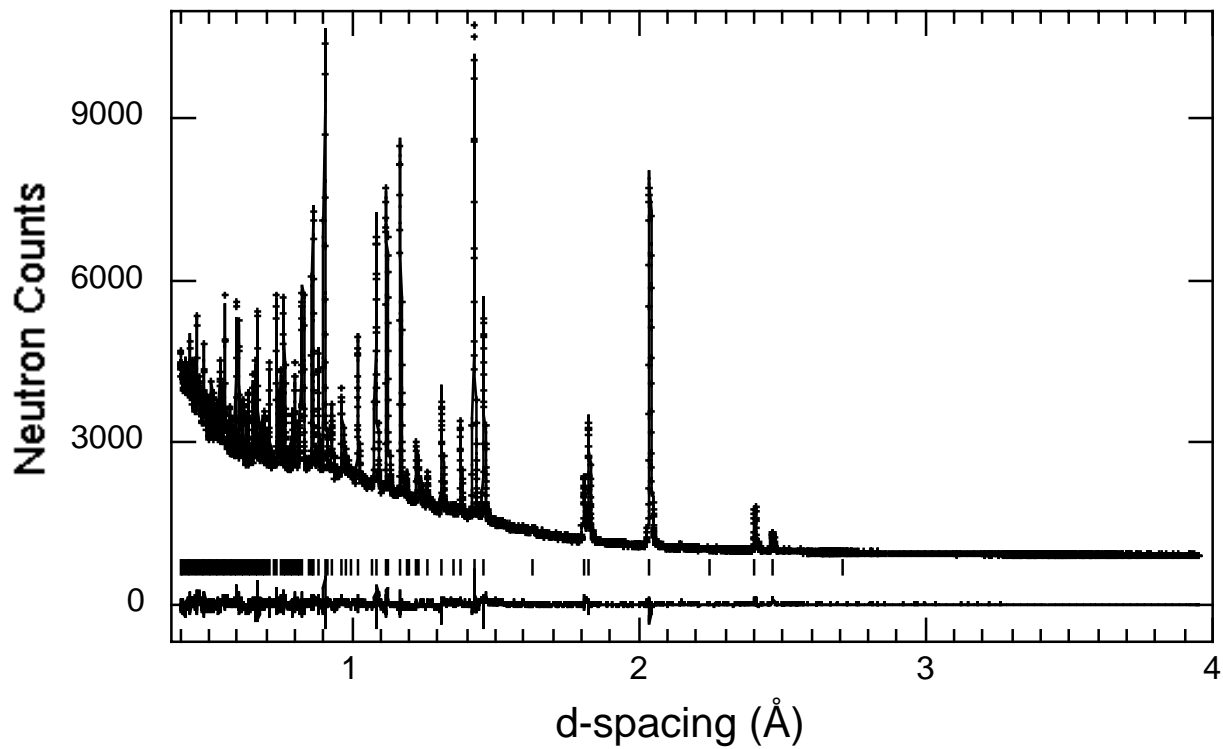

Fig. 3a

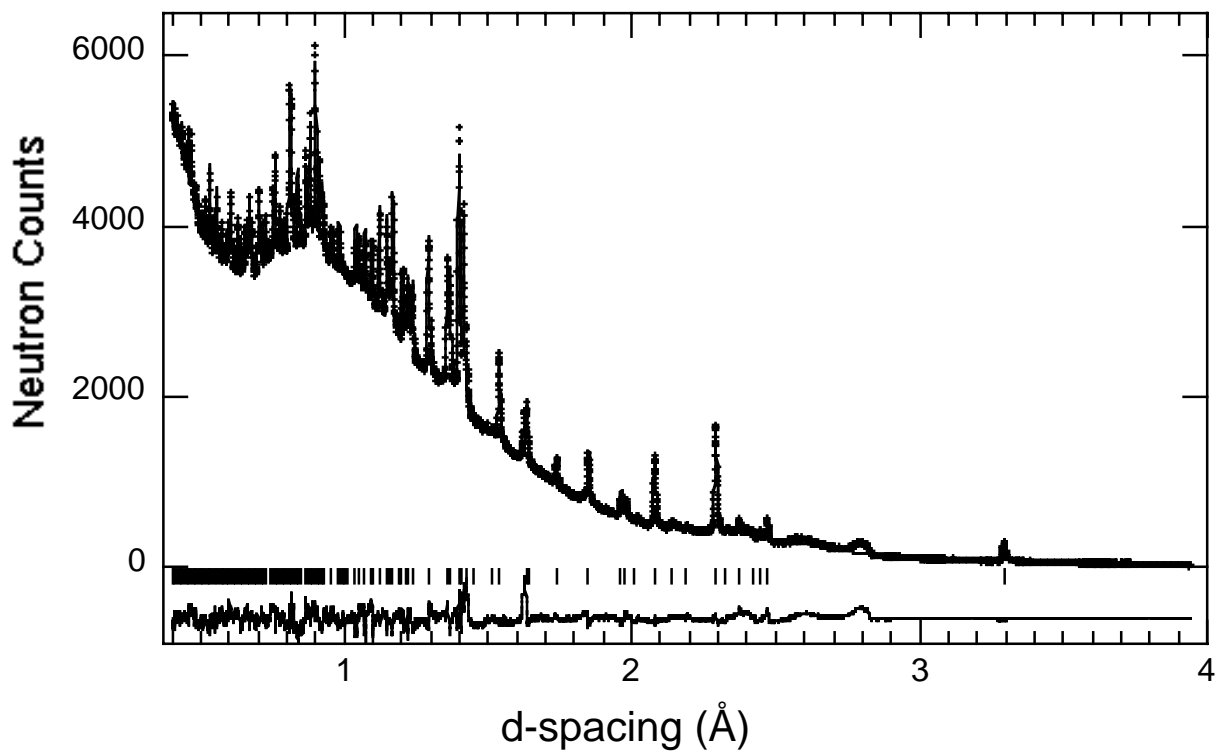

Fig. 3b

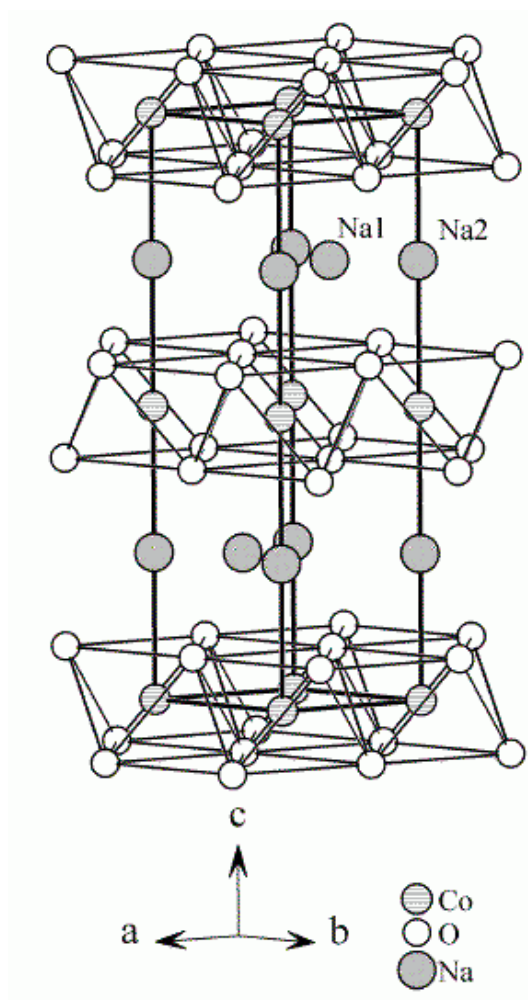

Fig. 4.

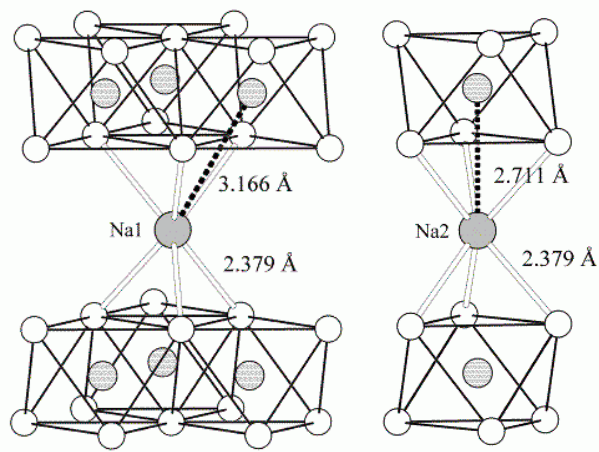

Fig. 5a.                           Fig. 5b.

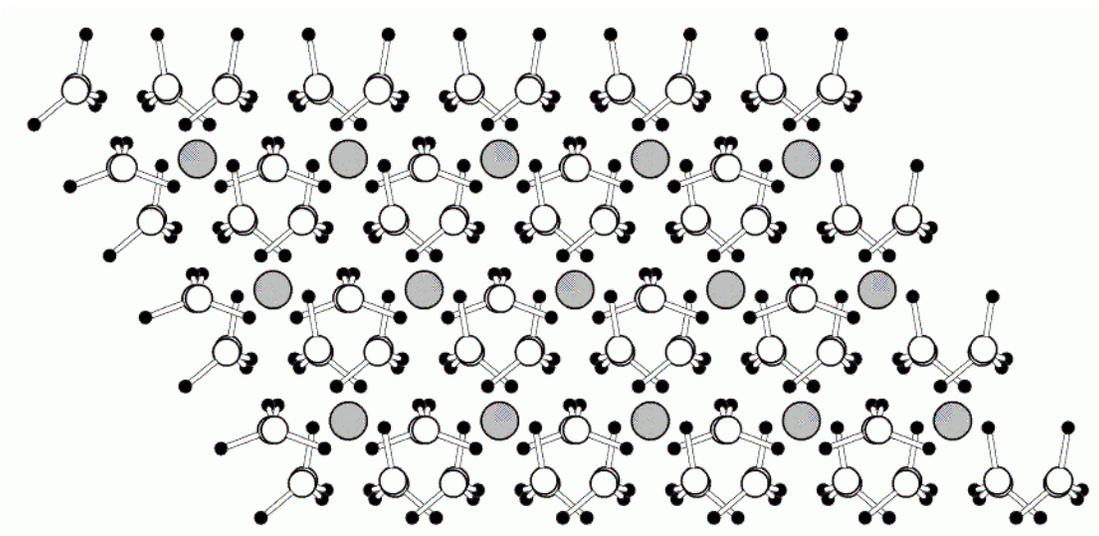

Fig. 6a

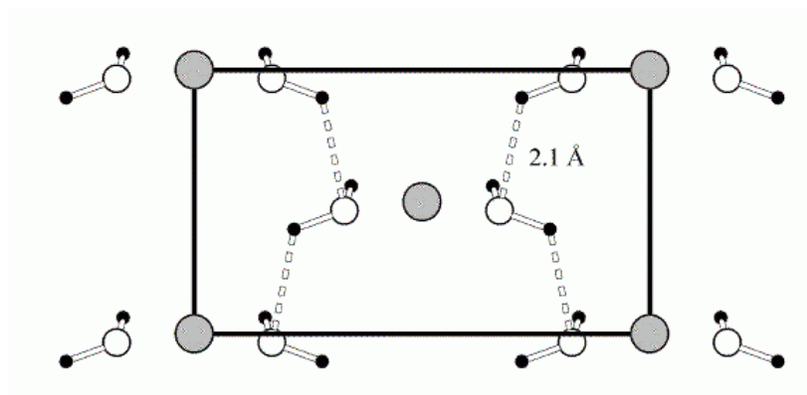

Fig. 6b

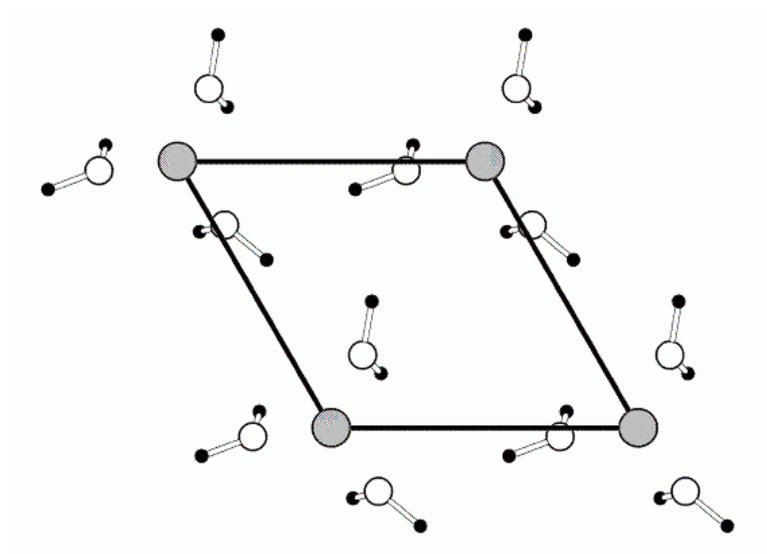

Fig. 6c.

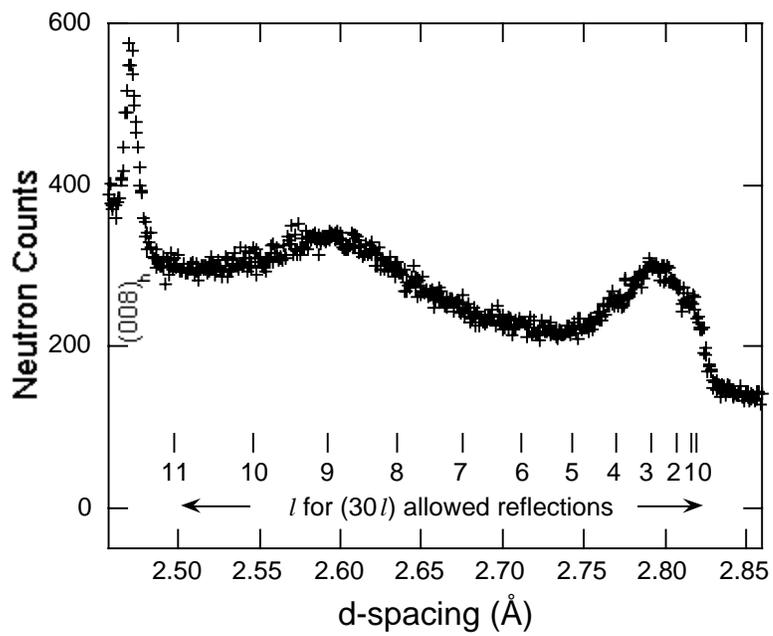

Fig. 7a

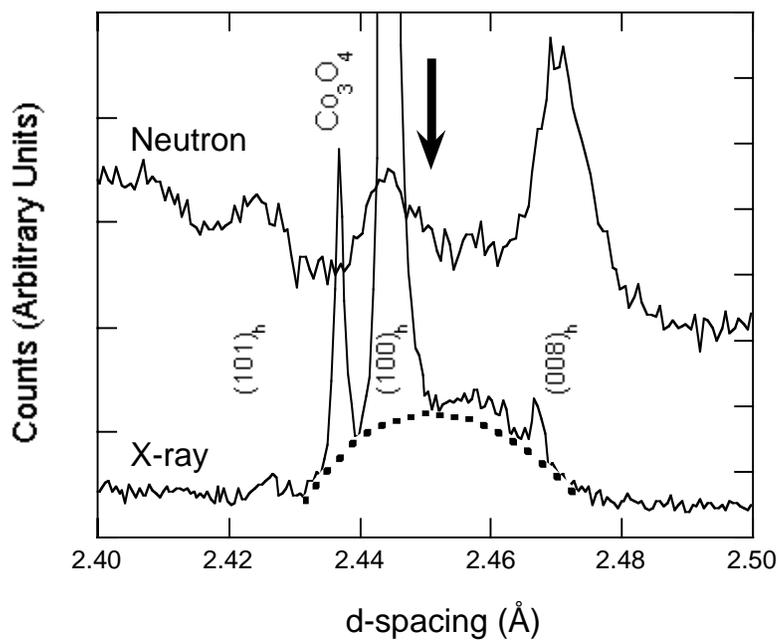

Fig. 7b

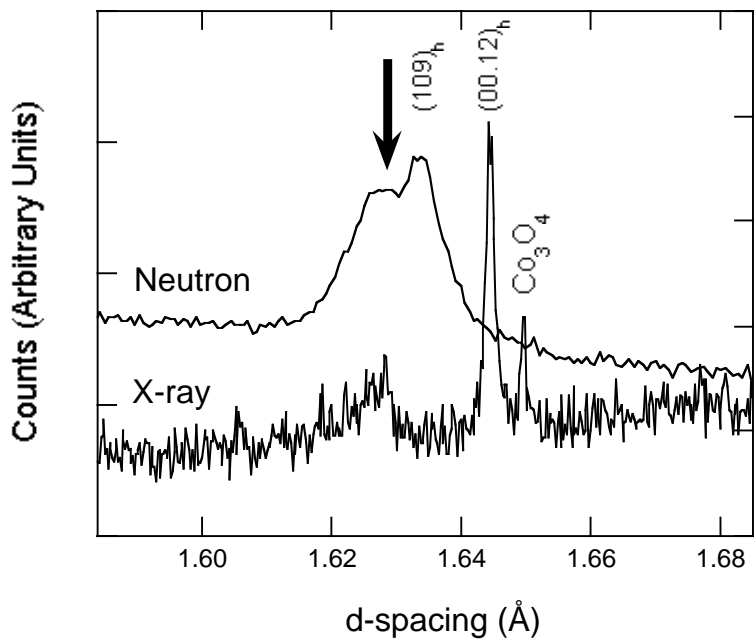

Fig. 7c.

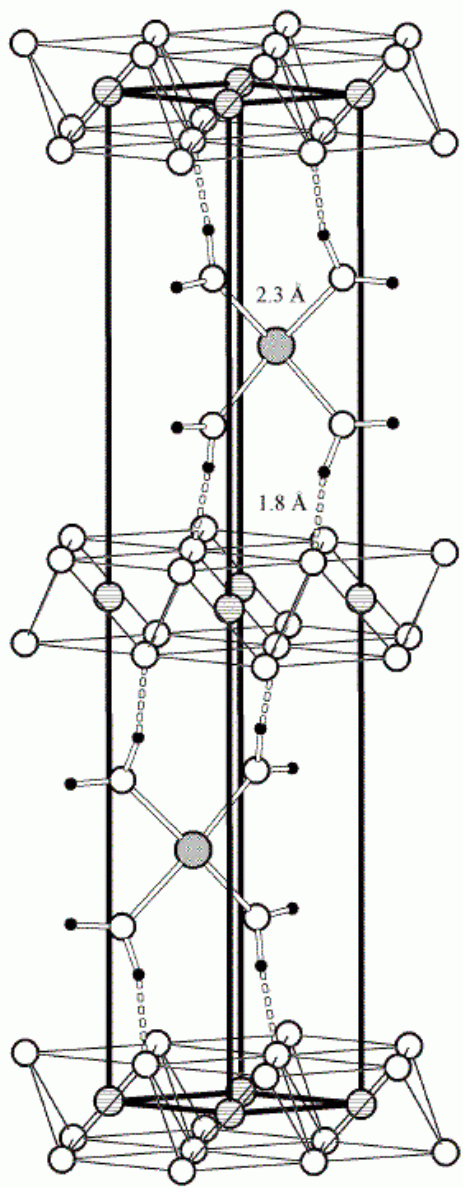 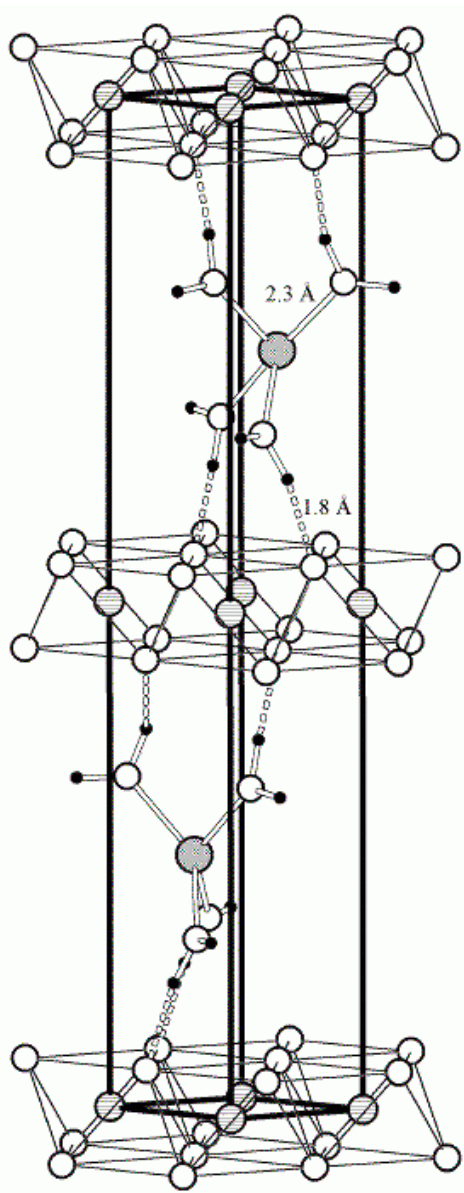

Fig. 8a                    Fig. 8b